\documentclass[acmsmall]{acmart}
\usepackage{amsmath,amsfonts}
\usepackage{algorithmic}
\usepackage{graphicx}
\graphicspath{ {./images/} }
\usepackage{textcomp}
\usepackage{xcolor}
\usepackage{tabularx}
\usepackage{datatool}
\usepackage{graphicx}
\usepackage{subfig}
\usepackage[T1]{fontenc}
\usepackage[utf8]{inputenc}
\AtBeginDocument{%
  \providecommand\BibTeX{{%
    \normalfont B\kern-0.5em{\scshape i\kern-0.25em b}\kern-0.8em\TeX}}}

\setcopyright{acmcopyright}
\copyrightyear{2022}
\acmYear{2022}

\acmConference[CHI '22 Workshop: HCI Across Borders]{CHI '22}{April 13, 2022}{New Orleans, LA}
\acmBooktitle{CHI’22 Workshop: HCI Across Borders, April 16, 2022, New Orleans, LA, USA, 15 pages}

\usepackage{appendix}
\usepackage{tabularx}
\usepackage{booktabs}
\usepackage{hyperref}
\usepackage{subcaption}

\usepackage{float} 
\usepackage{soul,color}
\usepackage[flushleft]{threeparttable}

\begin{document}

\title{An Initial Study Review of Designing a Technology Solution for Women in Technologically Deprived Areas or Low Resource Constraint Communities}

\author{Jones Yeboah}
\email{yeboahjs@mail.uc.edu}

\affiliation{%
  \institution{University of Cincinnati}
  \city{Cincinnati}
  \state{Ohio}
  \country{USA}
  \postcode{45221}
}
\author{Sophia Bampoh}
\email{BAMPOHS@ccf.org}

\affiliation{%
  \institution{Cleveland Clinic Florida}
  \city{Weston}
  \state{Florida}
  \country{USA}
  \postcode{33331 }
}
\author{Annu Sible Prabhakar}
\email{annu.prabhakar@uc.edu}
\affiliation{%
 \institution{University of Cincinnati}
 \city{Cincinnati, Ohio}
 \country{USA}
 \postcode{45221}}

\renewcommand{\shortauthors}{}
\begin{abstract}
In the West African country of Ghana, depression is a significant issue affecting a large number of women. Despite its importance, the issue received insufficient attention during the COVID-19 pandemic. In developed countries, mobile phones serve as a convenient medium for accessing health information and providers. However, in Ghana, women's access to mobile phones is limited by cultural, social, and financial constraints, hindering their ability to seek mental health information and support. While some women in deprived areas can afford feature phones, such as the Nokia 3310, the lack of advanced smartphone features further restricts their access to necessary health information. This paper reviews the potential of Unstructured Supplementary Service Data (USSD) technology to address these challenges. Unlike Short Messaging Service (SMS), USSD can facilitate data collection, complex transactions, and provide information access without the need for internet connectivity. This research proposes studying the use of USSD to improve access to mental health resources for resource-deprived women in Ghana.
\end{abstract}


\maketitle

\section{Introduction}
In Ghana, depression is a major issue affecting a significant number of women, yet it has not received adequate attention, particularly during the COVID-19 pandemic. While mobile phones are a common medium for accessing health information and providers in developed countries, Ghanaian women face barriers such as cultural, social, and financial constraints that limit their mobile phone access. This limitation adversely affects their ability to gather mental health information and seek support. Women in deprived areas may afford feature phones like the Nokia 3310, but these lack the advanced features of smartphones, further restricting access to health information online.

To address this problem, previous research, including the seminal paper "The Third Universal App," suggests using Unstructured Supplementary Service Data (USSD) technology. Unlike Short Messaging Service (SMS), USSD supports data collection, complex transactions, and information access \cite{b1}. USSD's session-based and screen-designed platform can provide universal access to various user classes \cite{b1}. This research aims to explore the use of USSD to enhance access to mental health resources for women in deprived areas of Ghana.

The study will focus on regions where women face extreme poverty, inadequate healthcare, no internet connection, and are influenced by cultural and religious beliefs. Internet access is limited and expensive, with coverage often not reaching these areas. USSD offers a solution, as it is free on all networks in Ghana and does not require internet connectivity.

\begin{figure}[H]
    \centerline{\includegraphics[width=70mm,scale=0.75]{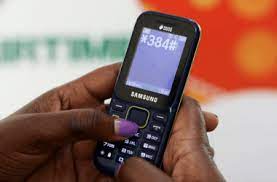}}
    \caption{A woman dialing a USSD short code on a feature phone}
    \label{fig:FeaturePhone}
\end{figure}

To access mental health information via USSD, users will dial a dedicated short code, prompting their feature phones to display information on common depression symptoms among women. They will follow text prompts, answer a series of questions, and provide responses based on their experiences. The system will record their mobile numbers, forwarding the responses to the nearest satellite health unit call center via SMS and email. Based on the information provided, they will be connected to the nearest satellite health unit for assistance. An example of a USSD is shown in \autoref{fig:FeaturePhone}

By leveraging USSD technology, this study aims to bridge the gap in mental health information access for resource-deprived women in Ghana, providing a practical solution to a critical health issue.
\section{Literature Review}
\subsection{Introduction}
According to the American Psychiatric Association \cite{b49}, approximately one out of every six women suffers from depression during their lifetime and one out of every fifteen adults suffers from depression every year. The Mayo Clinic has identified that postpartum depression symptoms can include \cite{b48} : experiencing depressive symptoms or mood swings, excessive crying, irritability, persistent tiredness, finding it difficult to bond with your newborn, feeling detached from closed friends, colleagues, relatives and family members, food aversion, insomnia, lack of interest in activities you previously enjoyed, insecurities about parenting, hopelessness, lack of confidence, and impaired concentration that can ultimately lead to deficiencies in decision making. This, in the end, can be complicated by anxiety and constant thoughts of death or suicide.

There is a high prevalence of mental illness globally \cite{b4}, which contributes to 14.3\% of all deaths world wide \cite{b5}. Most developing countries have significantly higher percentages of mental health issues, reaching over 90\% in some cases \cite{b20,b21}. The rise of mental illness in global development scenarios has been attributed to a psychological, social and biological factors. It has also been shown that human rights violations \cite{b7}, political oppression \cite{b8}, and regional ruralization \cite{b9} are  major risk factors of developing mental illness. Mental health challenges are mostly seen as direct roadblocks to successful achievement of developmental goals. Given its direct impact on the psychosocial development of human beings, mental illness disrupts the growth of functional societies and thus economies, which in turn disrupts critical structures for achieving development \cite{b18}. This vicious cycle ultimately worsens mental illnesses due to the stress of living in underdeveloped communities. In all mental illness and societal and economic expansion are inextricably linked \cite{b19}. Therefore, mental health has recently been incorporated into the Sustainable Development Goals \cite{b10,b11} due to its significant impact on a variety of issues, including access to education, healthcare, social cohesion, economic transformation and poverty alleviation. For this reason, networks have been scaled up to tackle the burden that mental health places on global development, with a key organisational unit being "The Movement for Global Mental Health" \cite{b12} which includes renowned institutions as the "World Health Organization (WHO)" \cite{b12}.

It was discovered during Covid-19 that mental health cases in developing countries with limited resources were not being adequately addressed. Based on research conducted during the pandemic, it was found that Covid-19 exacerbated mental health challenges with programs like governmental lockdown rules intermingling with fear and anxiety from increased death rates experienced worldwide. Considering the magnitude of the pandemic event and its effect on global mental health, it deserves special attention. This is more so crucial in the developing world where it is essential that policies are instituted to address challenges faced in both cities and more rural communities that tend to be resource limited \cite{b52}.

Recent changes made to guidelines for obstetric care, have highlighted that women who are in the peripartum  period are more likely to suffer from issues affecting their mental well being. This has been compounded by the pandemic, with Ajayi et al. \cite{b53} noting that various COVID-19 pandemic related stressors have significantly affected pregnant women as well as postpartum mothers in Africa. Indeed, mental health among African mothers worsened during the pandemic, due to the region's weak health system, ineffective policies, and lack of facilities dedicated to mental health matters. This is exacerbated by the fact that poverty rates are high and women lack reliable access to maternal care. Due to the heterogeneity within and between African regions, multifaceted strategies and interventions are required in order to combat mental illness, with special consideration being given to mental health in peripartum women and technological solutions in this paper.

 To address this apparent treatment gap, the fields of Information Technology and more specifically Human Computer Interaction, need to study and develop interventions that take into consideration complex and intersecting systemic constraints \cite{b22,b23,b24}. "Human-computer interaction" (HCI) \cite{b80}, is increasingly being adopted in mental health programs due to its high prevalence and impact on understanding human-technology interactions. According to Caroll \cite{b80}, "Human-computer interaction (HCI) is an area of research and practice that emerged in the early 1980s, initially as a specialty area in computer science embracing cognitive science and human factors engineering. HCI has since expanded rapidly and steadily for three decades, attracting professionals from many other disciplines and incorporating diverse concepts and approaches" \cite{b80}. This expansion has made it amenable to improving mental health access opportunities. Significant HCI research is being conducted in order to understand and predict how people express mental distress online \cite{b13,b14}. Therapies necessary for mental health treatment traditionally require complex require in-person interactions between patient, family and health workers which remains inaccessible to many people. Technological tools such as HCI utilize social engineering methodology to increase the reach of therapeutic approaches via technology based scale up of access to mental health care \cite{b16}. For example, HCI tools like “personal informatics systems”\cite{b15} engage patients to improve health care access. In maternal mental health care, self-centric engineering is a personal informatics tool that includes "self-tracking, self-reflection, self-knowledge, self-experimentation, self-improvement" \cite{b15}.  HCI tools like this improve access and delivery of mental health care and thus mediate recovery from mental illness \cite{b17}.

\subsection{Maternal Depression in Ghana and Africa}
Research has shown that women of reproductive age are more likely to suffer from depression \cite{b25} with a high confluence around the perinatal period. \cite{b81}. Notably, there is a difference in depression rates in the different stages of pregnancy as noted in a Lancet meta-analysis of low-middle income countries by Gelaye et al. \cite{b37}. Per the meta-analysis, the estimated rate of depression was 25.3\% vs 19.0\% in the antepartum vs the postpartum period respectively. There is also a notable difference in peripartum depression between the developed and developing world. Among mothers living in developed countries, depression is estimated to affect 10.4\% to 16.4\% of them \cite{b26,b81}, while depression rates are noted to be significantly higher and range from 14\% to 50\% in low income countries \cite{b82}. More specifically, in Africa depression affects 18\% of this category of women \cite{b27}. 

Ghana is a West African nation with a population of approximately 31 million per the latest census data in 2021 \cite{b84}. In Ghana, studies on peripartum women have shown depression prevalence rates ranging from 3.8\% to 37\% \cite{b28,b32,b33,b34,b35,b83}. Per a WHO report, only 1\% of Ghanaian patients with mental disorders (including peripartum depression) obtain any form of psychiatric healthcare \cite{b94}. The impact of depression is undeniable, and it contributes significantly to maternal and infant mortality and morbidity \cite{b36}. For peripartum women, multiple significant consequences of depression have been documented including increased risk of obstetric complications and preterm labor \cite{b89,b90,b91,b92}, economic losses of an inactive work force and disproportionately high suicide rates. A study in Mozambique revealed an alarming 30\% suicide rate in the peripartum period \cite{b93}. 
Multiple studies have also detailed how maternal peripartum depression affects infant psychosocial development \cite{b85}, intellectual competence \cite{b86}, emotional development and rates of psychiatric morbidity\cite{b86,b87,b40,b41}. In developing countries like Ghana, the WHO also notes that mental health problems among mothers are linked to low birth weight, decreased breast-feeding \cite{b42,b43,b44,b45}, stunted growth, severe malnutrition, higher rates of diarrhoeal illnesses and lower compliance with immunization schedules \cite{b82,b38,b39}.

These myriad impacts of maternal depression underscore the importance of research that aims at identifying and treating patients with the condition.

\subsection{USSD}
\subsubsection{\textbf{Introduction}} 

The Global System for Mobiles (GSM) network was built in order to provide a basic telephonic service for real-time oral communication. Due to the evolution of GSM technology, however, new means of communication have been developed including SMS technology that was initially introduced to assist subscribers to exchange text messages. In contrast, USSD was developed to provide real-time data communication between subscribers and supplementary services such as banking, insurance, and educational sectors in order to meet the needs of subscribers. Even though, USSD is fully penetrated into daily life in countries like Ghana, users are often unaware of how integrated it is into their lives. For example, loading airtime credit and checking airtime credit balance in Ghana has long been a USSD based service.

\subsection{Why USSD?}

There are numerous benefits associated with USSD technology \cite{b54}. USSD is cost efficient and it supports GSM networks at a low cost because it runs on the same protocol. It is also relatively easy to develop and integrate USSD applications due to the fact that the protocol is not complicated, thus reducing marketing costs for developers and companies. USSD is also highly customizable, making it suitable for a wide range of governmental and private sector institutional needs. USSD provides a quick response time and this real-time interaction with services is a big draw for subscribers in comparison to SMS where there is a delay while awaiting response. Additionally, USSD is highly interactive due to its session-based nature, thus operators can create interactive applications like mobile banking and merchant solutions. In the context of cellular networks, USSD technology facilitates real-time communication between mobile phones and application servers through session-based messaging\cite{b3}. As a result of USSD service, the mobile network is capable of supporting a wide variety of portable handheld devices, including basic feature phones as well as smartphones. USSD has traditionally dominated the financial technology space in rural and urban communities in Africa, mainly for mobile money transactions \cite{b3}.

Due to USSD technology's independence from the internet, health applications can be developed around it in resource limited settings. In resource limited environments, internet service is constrained by poor network infrastructure, high costs, and limited access by other users. Most people cannot afford to stay online 24 hours a day, even with smartphones like those made by G-Tide and Huawei.
In the health sector, USSD technology can be use to provide interactive electronic health records (EHR) systems for healthcare workers and patients to check health information and receive notifications via SMS using USSD codes. This real time health information (EHR data) can also be integrated across multiple healthcare facilities in different regions and stored it in a repository that can be accessed from anywhere\cite{b2}. In this way USSD can provide healthcare workers and patients  real-time access to health information from a stored repository and allow for well-informed decisions and treatment strategies to be instituted to improve not only access but also the quality of health care \cite{b3}.

\subsection{Design Perception of USSD System}
There is a parallel components that is the core design concept of USSD. In parallel USSD design, each component is independent and can function on its own but maintains a flexible connection to other components. Therefore, USSD works in a fashion akin to a plug-and-play system. This leads to the following special properties of each component. Firstly, components can be implemented on different nodes with little additional effort. For instance, by introducing a micro-service architecture. Secondly, The USSD system allows for flexible connections between components and integration with external Application Programming Interfaces (API).
Thirdly, USSD is capable of processing transactions and activities on the platform synchronously and asynchronously, as well as being able to support both scenarios. In addition, it is imperative for a USSD system to have the capability to work on multiple system platforms at the same time, which gives it the ability to function in parallel with several other systems at once. It is through this approach that the system is able to increase the application throughput, provide high responsiveness for inputs and outputs, as well as a more appropriate program structure. Fourthly, USSD has a fail-safe error recovery, such that in the event of service failure the system informs the related process and restarts or crashes silently if necessary. This leads to a high-availability system. Lastly, USSD has defined timeout values in each process. Thus, if one process fails, it will inform the other processes or crash itself. If it does not receive the desired response from the USSD gateway with its defined value as a result of not receiving a response from the USSD gateway, then it will fail. It is thus possible to obtain a mechanism for automatic garbage collection.

\subsection{USSD Technology and Its Applications in Ghana}
There is a growing telecom industry in Ghana, with a wide range of innovative technologies that are revolutionizing the telecommunication sector. MTN, Vodafone, and AirtelTigo are all mobile telecommunications companies (telcos) that use USSD codes to provide various services to their network subscribers. Using USSD services, users have been able to purchase air time, check airtime balances, purchase internet data bundles, send call me back requests, and receive notifications regarding insufficient credit. USSD shortcodes are also extensively used in the banking industry to access an interactive menu and conduct online banking activities, such as money transfers, account balance checks, utility bill payments, school fees payment, and monthly statement notifications\cite{b56}. Baffour \cite{b56} asserts that banking service providers continue to introduce mobile based USSD innovations in order to attract new customers and to increase their competitive advantage. In Ghana, telcos have formed alliances with various financial technology companies to extend the capabilities of USSD to offer services over mobile networks. Numerous projects have made significant progress in providing financial and real time services in the agricultural sector, education sector, and insurance sector. 

Ghana is a heavily agrarian society with major portion of the population engaged in farming. In Ghana, farms is mainly located in rural areas with limited technological resources. USSD has made it possible for farmers to sell their products to the government and receive payment via mobile money payment services, thus streamlining the supply chain process. As an example, a USSD product called MAgric is used by License Buying Companies (LBCs), District Managers (DM), Purchase Clerks (PCs) and farmers. In Ghana, the LBC has the responsibility of ensuring quality standards of cocoa, a major Ghanaian export. LBCs delegate to their DMs and PCs the role of direct from farm quality assurance, sealing and shipment to the LBCS which are primarily located in urban areas. This chain of middlemen  lengthened transaction time for both the supplying farmers and purchasing LBCs. The advent of USSD technology has led to speed and efficiency in these transactions. Now farmers are able to find ready markets and receive payments faster, while LBCs are able to track all activities that involve DMs, PCs, and farmers in rural areas without internet access. The use of this USSD based approach has greatly enhanced the effectiveness of monitoring and payments, thus enabling the LBCs, DMs, PCs, and farmers to communicate effectively \cite{b62}.

In the education sector, a flagship USSD product, "School Fees," which is available on a number of mobile platforms in Ghana. This USSD system allows parents living in low-resource communities without access to the internet to remotely make school fee payments for their children in all levels of education. Parents can also access other USSD educational services to track their wards real-time academic performance during a term. By doing so, parents and teachers have been able to exchange information and engage with each other effectively \cite{b63}.

During the Covid-19 pandemic, most insurance companies in Ghana migrated to digital solutions with USSD as the preferred technology. Prior to this period insurance policies were inequitably distributed with a high percentages of policies being held in urban areas. USSD technology allowed for ease of marketing and purchasing of insurance policies remotely and thus made the insurance market more equitable. In the comfort of their homes, customers can now purchase insurance policies for themselves or someone else, view insurance statements, check balance, pay premiums, make requests and withdraw funds by dialing a short code. Technology of this type promotes inclusiveness and makes it accessible to everyone.

The aforementioned examples shows the integration of USSD technology into the Ghanaian social fabric. Healthcare remains an untouched frontier in which USSD technology is yet to be explored in Ghana to improve quality health care delivery for the Ghanaian populace.

\subsection{Intersection of USSD Technology and Health Projects}
As a result of depression's prevalence and impact, public health has prioritized prevention and treatment. By introducing technological and innovative ideas for easy access to health information and for providing quality healthcare, digitization is one of the best approaches to revamping the health space. There have been a number of health technologies introduced to assist in the monitoring and assessment of health, which even extends to providing support services for people suffering from depression. The systems design and people’s everyday technology interactions and information gathered in electronic health records (EHRs), are providing huge growth in information on people’s health and behavior.

According to Amoakoh et al. \cite{b2}, USSD technology is necessary for providing advice to clinicians for the purpose of improving the outcome of clinical decisions in low-resource settings in developing countries such as Ghana. In their discussion, the authors discuss an innovative project that uses SMS as well as USSD to reach out to a variety of health facilities. Aiming to provide quality health care to the population, it was designed to enable health workers to access information readily to make informed decisions to improve maternal health. 

Zhou et al. \cite{b3} explore how USSD Technology is a cost-effective option for enhancing the work process of various health providers in low resource settings. As the platform is not dependent on the internet, users are able to view patient profiles, receive SMS notifications, and access real-time data, regardless of where they are located. The authors explains how effective the tool was able to bridge the communication gap between healthcare providers and patients for the purpose of providing the best possible care to patients. In addition, the authors propose a number of innovative features that can also be implemented using USSD technology. By providing self-help capabilities within the platform, patients will be able to access health information tips on their own and have access to USSD support services to assist with any questions they may have. Therefore, individuals will have easier access to health information tips on their own. A well-designed education and awareness program, readily available when needed, is critical to the health industry's transformation agenda.

Joseph et al. \cite{b46}, proposes the use of USSD technology to support transmission of patient data from caregivers at home to health facilities in the area with a focus on providing a better healthcare delivery system in South Africa. The authors also propose the use of USSD technology as a means of providing a patient monitoring system to assist medical staff to monitor patients' vital signs on a regular basis as part of the monitoring program. As a result of this system, home-based care workers are able to send patient information to a local clinic or hospital for processing. Nurses and doctors use the information on a desktop computer to speed up the decision-making process for patients and improve their health.

Judith also proposes USSD as a technology intervention tool for multi-cultural and multilingual populations in Ghana. The author discussed how USSD technology facilitates two-way communication that involve the Service Providers and mobile subscribers without the need for an internet connection. The paper also discussed the USSD benefit such as being able to obtain relevant health information, particularly prevention information by dialing an approved USSD short code via mobile phone. Additionally, it helped to bridge linguistic gaps in accessing health information, fostering relationships and promoting health in multicultural settings and communities \cite{b47}. 

Killian et al. \cite{b50} proposes the use of USSD Technology as a method for providing support for health workers in one of their studies, which focused on developing countries with low resource settings. In order to achieve effective data distribution, a feasible, scalable, and accurate system was developed. In Malawi, a developing country, the study used mobile phones to collect information regarding the quantity of samples collected from patients as well as the location of where to access the test report in order to deliver it to the diagnostic network. In the study results, the authors conclude that USSD-based systems can be a potentially viable, adaptable, effective solution to the problem of inaccurate, unresponsive diagnostic data that can be addressed in a feasible, effective, and adaptable manner. Furthermore, USSD has cultivated a high level of patronage among its users due to its ease of use and affordability. It was viewed as an effective tool for improving accuracy rates and does not impose a financial burden on users.

According to Nakibuuka et al. \cite{b51}, paper-based and SMS-based approaches pose a great challenge to the delivery of healthcare decision making processes in a resource limited settings in Africa due to the poor data collection and reporting processes. Among the challenges encountered was the poor quality of the report and no internet. In a study conducted by the authors, it was determined that USSD technology was the most effective method for producing quality reports with fast response time. The results of this study demonstrate that USSD can be used to improve accessibility and availability of healthcare data, maintain accuracy, and provide fast response times regardless of location. Through this approach, the government and the Ministry of Health are able to make informed decisions based on complete data across all districts in order to carry out the mandate of providing quality healthcare to all citizens. 

Umar et al. \cite{b55} explores USSD technology as a tool or means of facilitating development of an effective blood bank service in low resource settings in deprive communities. In order to increase patient accessibility, volunteers, replacement donors (family and friends) or compensated donors will be more readily accessible to patients. As a result of unsafe blood transfusions, which can lead to deadly infections and consequently death in needy and deprived areas, the authors discuss some of the health challenges these areas face. It is suggested by the authors that with the full implementation of the USSD, service availability will increase regardless of whether you reside in an urban or rural community with low resources. In developing countries in Africa, patients will be protected from the intricacies of unsafe blood and at a reasonable cost with improved blood transfusion services.

\subsection{Techniques and Initiatives for Design in Low Resource Settings}
Culture, resource limitations, and depression issues should not be underestimated when designing tools to help women in lower socioeconomic communities. In design considerations for the software product, Tuli et al \cite{b57} propose taking into consideration not only the individual affected by the depression issue, but also the whole family. It was discovered, through several discussions with patients and caregivers, that caregivers desire an approach to bridge the gap between caregivers and patients, such that it meets the acceptable standards in the medical field. According to the paper, the solution would be to develop two applications, one of which would be accessible by the woman who is experiencing depression and the other would be accessible by the caregiver. The study by Mehrotra et al. \cite{b58} explains how cultural constraints and uneasiness have made it difficult for women suffering from depression to communicate their problems to other family members. Throughout the paper, emphasis is placed on designing a solution that can address the needs of such a group of women in that circumstance. Introducing a design module that will incorporate support services, which will provide users with helpful tips and assistance as needed. When designing solutions for the women, it is extremely important to consider their emotional and psychological needs.

There is a high risk of accidental harm when working with communities in rural areas with low resources. The sustainability of interventions is of the utmost importance, because if access to care suddenly is cut off, it could be disheartening. In the absence of treatment, depression can develop into a chronic condition that affects the individual for a long period of time. Since past experiments on mental well being lacked a clear communication of aims \cite{b59}, interventions need to ensure that the methods used for analyzing data and protecting personal information are clearly communicated to those who are involved. In light of the fact that some types of technology may exacerbate depression symptoms, this is particularly important \cite{b60}. Collaboration with other researchers and experts is essential to agreeing on a design module. Thus, the work adheres to the general ethical framework for human and study research within the country in which it is conducted \cite{b61}.

\subsection{Instrument Of Measurement in Depression}

Multiple screening measures for peripartum depression exist including The Edinburgh Postnatal Depression Scale (EPDS) which has been extensively validated in the high income countries \cite{b95}, but also in Sub-Saharan African countries like Nigeria \cite{b96}, and South Africa \cite{b97}. Other screening assessments include the Patient Health Questionnaire (PHQ-9) \cite{b68}, and the Self-Report Questionnaire (SRQ-20), \cite{b69}.  The PHQ-9 is a validated self-administered depression screening tool used in European and American healthcare setups with high sensitivity and specificity \cite{b98,b99}. It has also been used with impressive results in Nigeria \cite{b100} and  for peripartum depression screening in Ghana \cite{b34} including in more rural settings. The World Health Organization has also developed The SRQ-20  tool to accommodate a wide range of sociocultural contexts with special focus on low literacy rates as is found in developing countries such as Ethiopia \cite{b29}, Pakistan \cite{b30}, and Nepal \cite{b31}.

In Ghana formal validation of screening measures for peripartum depression was investigated by Woebong et al in 2009 \cite{b34}. In a comparative assessment of EPDS, PHQ-9 and SRQ-20, it was shown that the PHQ-9 had higher validity, reliability, and superiority. This short 9-question survey tool with is ease of administration was easily translatable to a population that is largely illiterate postpartum Ghanaian women in Kintampo, which is a resource limited community in rural Ghana.

\subsection{Study Methods In Depression Research}

Online surveys remain one of the preferred methods of gathering data for studies in the medical and health research fields \cite{b71,b72}. It is widely acknowledged that since the development of online survey creation programs like Survey Monkey, researchers and participants now have an easy-to-use option for collating data, which has been beneficial for both parties. Consent is a cornerstone of ethical research. Many online survey design tools utilize branching logic and question layout to obtain consent prior to allowing survey questions to be displayed to participants  and to streamline the survey path based on their responses to prior questions. In surveys where data points are mostly qualitative, limiting the number of characters on responses given by research subjects allows you to condense information and avoid lengthy and unmanageable input. Other data collection methods include sliding scales, multiple choice questions, or both. However, careful consideration should be given to question design in these more analog approaches as misconstrued questions may lead to incorrect or partially correct responses that may skew the results. Pre-testing is a valuable validation tool that is often used in the refinement an online survey's design. In the end, when the electronic data or output is finally collated  from an online survey it can be more efficiently analyzed with back-end tools and often at a more affordable price \cite{b71,b73,b74}.
Overall, a survey conducted online offers several advantages, including the ability to increase research subjects' willingness and capacity to participate anonymously and voluntarily and thus achieving a higher response rate in an increasingly technologically inclined and aware populace. The use of online research methods have been shown to improve the sensitive and thus reliability of results \cite{b76,b77}.

In my proposed study, the limited availability of internet infrastructure precludes or severely limits the use of online surveys for data collection. I therefore propose the use of a USSD based approach for data collection system using USSD, which is accessible to all people regardless of internet access.
Indeed, Nakibuuka et al. \cite{b51} developed a user-centred methodology for their research on collecting Ministry of Health (MoH) data in Uganda to facilitate health decision making at the district and national level. They noted poor quality of reported data due to the substandard modalities used in the collection and transmission of data, including paper-based and SMS approaches. To improve data collection, the researchers developed and implemented a "USSD-based health data reporting intervention in a district in Uganda" \cite{b51}. Their study showed that a USSD-based system for reporting ultimately resulted in complete, accurate and timely reporting in the resource-limited healthcare settings of Uganda. However, poor network coverage impeded reporting and infrastructural improvements a required to scale USSD systems.

Reviewing the literature, demonstrates that an HCI approach which employs user-centered design works iteratively and integrally to satisfy the needs of the user and deliver the solutions that clients expect. End users are engaged at every phase of the development process using tools such as interviewing, conducting surveys, and occasionally conducting brainstorming sessions in order to understand the users' needs \cite{b78}. HCI engineering expert Don Norman highlighted the importance of improving the experience of end users based on their experience on the product. \cite{b79}. An integration of the end user in the design process will help improve user experience of the USSD product in measuring depression. For the purpose of my proposed project, stakeholders to be engaged include healthcare providers, patients and care givers in an effort to develop a user friendly USSD product that will lead to increase participation of end-users and thus enhance data collection.

According to Haiduwa et al. \cite{b63}, their study of the use of USSD to enhance parent-teacher engagement was conducted using both qualitative and quantitative methodology approaches with an experimental design. The main source of information was provided by online surveys and virtual interviews. After reviewing the study, I find it useful to utilize the experience-centered design method, so that we may utilize the experience of pregnant women and health care providers to improve the new USSD solution based on their expectations and concerns. The idea of enriching the user experience of the stakeholders should be incorporated into the design of the solution. This study will involve the development of scenarios and use cases, as well as providing a great deal of information regarding the methods by which users are able to achieve their goals.

Umar et al. \cite{b55} conducted a study where they utilized USSD technology in order to connect to a centralized database so that patients could have easy access to donors when it comes to challenging their health. The approach method for this study focuses on using a model driven approach as the basis for the application to be developed in order to maximize the safety of blood transfusions for patients. In the review of the literature, it was determined that very little attention had been given to the use of this approach in designing solutions, based on the study results. The approach does, however, provide solutions to describe key stakeholders in the development process, such as patients, donors, and hospitals, as integral parts of the solution. The results of the analysis indicated that building solutions of this kind is both convenient and quick as it makes the development process easier.

The Model-driven Development (MDD) process allows software to be written and implemented quickly, efficiently, and with the least amount of cost as possible. MDD is the ability in creating a model of a software application. The model specifies how the system is to operate and tested with Model-based testing (MBT) after it is created, and then it can be deployed once the testing is complete. As an example, this approach can be used in our study of USSD in identifying the key entities or actors involved, such as patients, health providers, caregivers, and defining the specific attributes and functionalities during the development phase. By adopting this method approach, you will be able to reap a number of benefits that come along with it. Before coding begins on a software product, the behavior or action of the product is defined as what it should be. To define the functionality of each service, the software engineers along with other team members collaborate and work together in order to define the requirements. In terms of testing and deployment, MDD is a superior approach to other traditional development approaches. In the Software Development Life Cycle (SDLC), MDD is often used in conjunction with agile software development methods to develop software. Agile development breaks down projects into changing sprints of activities throughout each section. Iterations that are classified as short are acceptable during Agile model-driven development (AMDD), while the model is constantly revised. In AMDD, modeling is done in both sprints and coding is done in subsequent sprints. Finally, this approach will be useful in the development of the USSD system \cite{b70}.

\subsection{Research Gaps}
Review of the research efforts suggests that a proposed project that seeks to improve identification and access to mental health resources for peripartum mothers in Ghana is lacking. Efforts have been made to improve service delivery (including healthcare) in rural communities despite a lack of internet infrastructure. This has included the use of USSD based technology with some good effects seen. However, literacy rates and ability to read is poor in the rural regions of Sub-Saharan Africa and the USSD applications so far used have been in the official language which are often European ( eg English and French). So far, efforts to bridge the language barrier that patients must surmount are lacking. A USSD application developed in a local language can help bridge this gap and improve the technologies uptake in a study population. The often multilingual nature of African countries with many sub-dialects may proof to be a hindrance, but I propose concentration of the predominant local language in the region. For example, while there are over 49 languages in Ghana, most people speak one of four languages Twi, Ga, Ewe, or Dagbani. Therefore, a focus on one of these depending on the geographic location targeted would improve penetration and acceptance of the technology.

Another glaring dilemma so far not addressed are the security vulnerabilities associated with USSD technology being unencrypted. It is critical that a patients privacy and integrity are respected. Furthermore, in most jurisdictions, the patient has autonomy over their data and are entitled to it and they alone are allowed to determine who has access to their protected healthcare information. 
An unencrypted USSD based application does not take into consideration the ethical transfer of information. This review shows a lack of data on how to incorporate security mechanisms to protect patient health information. One such method that I propose would be two factor authentication that will require users to enter a pin prior to proceeding to view the requested information. This security feature like the language options feature require further detailed research.  We will describe the USSD system and its technical specifications in the next sections.

\subsection{Conclusion}
In conclusion, this initial review provides an overview of USSD technology, highlighting its features, capabilities, and potential to address the issues at hand. Additionally, it identifies gaps and areas for improvement, paving the way for continued research in this field as we move to the next phase of the research study.

\end{document}